\documentclass[11pt]{article}

\usepackage[a4paper,margin=1in]{geometry}
\usepackage{amsmath,amssymb,amsfonts}
\usepackage{physics}
\usepackage{bm}
\usepackage{booktabs}
\usepackage{setspace}
\usepackage{titlesec}
\usepackage{graphicx}
\usepackage{caption}
\usepackage{float}
\usepackage{placeins}
\usepackage{hyperref}

\title{Entropy production reveals hidden dynamical constraints rather than stochastic disorder}
\author{Patrick Romanescu\\
\texttt{patrick.romanescu@queensu.ca}}
\date{}

\begin{document}

\maketitle

\section*{Abstract}

Entropy production is often interpreted as a proxy for microscopic disorder or environmental roughness in stochastic systems. We test this interpretation using controlled simulations of overdamped stochastic dynamics on curved surfaces in which local noise, geometry, and forces are held fixed while global constraints are varied. Trajectories are generated for particles evolving toward a central attractor, and entropy production is quantified using both a continuum probability-current estimator and coarse-grained Markov transition statistics across multiple spatial and temporal resolutions. Across systematic sweeps of timestep size, domain extent, and boundary topology, entropy production is governed primarily by constraint-induced probability flow rather than local stochastic variability. Periodic domains that permit sustained circulation yield substantially higher entropy production than reflecting domains despite identical local stochastic structure, with the magnitude of the separation depending on domain extent. In contrast, coarse-grained estimates decrease as temporal resolution increases and rise with finer spatial binning, demonstrating that discrete estimates depend strongly on observation scale and may fail to resolve topology-induced irreversible structure. Ergo, entropy production is not a direct measure of environmental roughness or randomness. Instead, it quantifies how strongly system dynamics are driven away from reversibility by global constraints, geometry, and the space of allowed trajectories. Interpreted in this way, entropy production maps function as diagnostics of organized probability flow and provide a principled method for detecting hidden dynamical constraints from trajectory data alone.

\section{Introduction}

Understanding how stochastic systems evolve under hidden constraints is a central problem across physics, biology, and complex systems science. In many real-world settings one does not directly observe forces, potentials, or geometric structure, but only trajectories \cite{ciliberto_experiments_2017}. From intracellular transport and molecular motors to atmospheric flows and financial markets, the inference challenge is therefore inverse: given path data alone, what properties of the underlying dynamics can be recovered? Classical approaches estimate drift and diffusion from observed increments, thereby reconstructing effective forces or noise amplitudes \cite{pavliotis_stochastic_2014,risken_fokker-planck_1996}. While powerful, such methods primarily characterize variability and deterministic tendencies; they do not directly identify where a system dissipates energy or breaks time-reversal symmetry. Yet irreversibility, rather than noise magnitude, is the defining signature of nonequilibrium structure \cite{seifert_entropy_2005,seifert_stochastic_2012}.

Stochastic thermodynamics formalizes this intuition by extending classical thermodynamic notions, such as work, heat, and entropy production, to individual trajectories and to well-defined non-equilibrium ensembles \cite{seifert_stochastic_2012}. Within this framework, entropy production provides a quantitative measure of time-reversal symmetry breaking and dissipation in stochastic dynamics \cite{spinney_entropy_2012,van_den_broeck_three_2010}. Global entropy production rates distinguish equilibrium from nonequilibrium regimes, constrain fluctuations via fluctuation relations, and quantify dissipation into hidden degrees of freedom \cite{jarzynski_nonequilibrium_1997,crooks_entropy_1999,lebowitz_gallavotticohen-type_1999,bertini_macroscopic_2015}. However, most commonly used entropy production summaries are spatially averaged or trajectory-integrated, yielding a single scalar for an entire process. Such global metrics cannot localize where dissipation occurs, nor can they distinguish whether irreversibility arises from geometry, boundary constraints, coarse graining, or external driving. For systems evolving on heterogeneous substrates or curved manifolds, this limitation is particularly severe \cite{hsu_stochastic_2002,ikeda_stochastic_1981}.

The present work addresses this gap by formulating and testing a local entropy production density that can be estimated from trajectory statistics. For diffusion processes, deviations from detailed balance are encoded in the probability current field, and a quadratic functional of the current against the inverse diffusion tensor yields a scalar density with a direct interpretation as local irreversibility \cite{lebowitz_gallavotticohen-type_1999,van_den_broeck_three_2010,spinney_entropy_2012}. Regions where this density vanishes are locally time-reversible; regions where it is elevated identify locations where ordered probability transport is converted into dissipation. In this sense, entropy production functions not merely as thermodynamic bookkeeping, but as a spatial probe of hidden dynamical structure \cite{gingrich_dissipation_2016,nardini_entropy_2017}.

A persistent intuition nevertheless remains: that entropy production should serve as a proxy for microscopic disorder, environmental roughness, or randomness. To test whether this interpretation is justified, we construct a controlled model system: overdamped stochastic motion on a curved surface under a confining potential \cite{risken_fokker-planck_1996,pavliotis_stochastic_2014}. This system allows geometry, noise, topology, and boundary conditions to be manipulated independently while trajectories remain fully observable. The critical intervention is that local ingredients (noise amplitude, drift structure in the interior, and surface geometry) are held constant, while only global constraints (boundary topology and domain extent) are varied. Using both a continuum probability-current estimator and a coarse-grained Markov estimator rooted in time-reversal asymmetry of transitions \cite{schnakenberg_network_1976,roldan_entropy_2012,norris_markov_2009}, we evaluate entropy production across resolutions and constraint regimes, and we interpret the resulting scale dependence in terms of coarse graining and effective thermodynamic descriptions \cite{esposito_stochastic_2012,polettini_effective_2017}.

The results support a general reinterpretation: entropy production primarily tracks organized, constraint-permitted probability flow rather than raw randomness. In particular, topology that permits sustained circulation amplifies entropy production even when local stochastic structure is unchanged, while coarse-grained estimates depend strongly on the observer’s spatial and temporal resolution \cite{esposito_stochastic_2012,polettini_effective_2017}. These findings position entropy production maps as diagnostics of constrained flow and suggest experimental strategies for identifying hidden dynamical constraints from trajectory data alone, complementing broader perspectives that link irreversibility to information-theoretic asymmetry and dissipation \cite{kawai_dissipation_2007,parrondo_thermodynamics_2015,cover_elements_2005}. To our knowledge, this is the first controlled demonstration that entropy production is governed primarily by global dynamical constraints rather than local stochastic structure.

\section{Theory}

\subsection{Embedded surface geometry and induced metric}

We consider an embedded two-dimensional surface $\mathcal{M}$ embedded in three-dimensional Euclidean space as the graph of a smooth height function
\begin{equation}
\mathcal{M}=\{\left(x,y,z\right)\colon\ z=f\left(x,y\right)\},\qquad\qquad \left(x,y\right)\in\Omega\subset\mathbb{R}^2,
\end{equation}
with $f:\mathbb{R}^2\rightarrow\mathbb{R}$ defined by a superposition of a quadratic bowl, two localized Gaussian deformations, and a sinusoidal ripple:
\begin{equation}
\begin{aligned}
f\left(x,y\right)
&=-a\left(x^2+y^2\right)
+0.22\,\exp\left(-1.8\left(\left(x-0.7\right)^2+\left(y+0.25\right)^2\right)\right)\\
&\qquad
-0.18\,\exp\left(-2.2\left(\left(x+0.9\right)^2+\left(y-0.15\right)^2\right)\right)
+0.06\,\sin\left(5.2x\right)\cos\left(4.6y\right),
\end{aligned}
\end{equation}
with $a=0.35$. Let $\nabla f\left(x,y\right)=\left(\partial_x f,\partial_y f\right)^\top$. Using the parameterization $\phi\left(x,y\right)=\left(x,y,f\left(x,y\right)\right)$, the induced Riemannian metric on $\Omega$ is
\begin{equation}
g\left(x,y\right)
= D\phi\left(x,y\right)^\top D\phi\left(x,y\right)
= I_2+\nabla f\left(x,y\right)\,\nabla f\left(x,y\right)^\top,
\end{equation}
a $2\times2$ symmetric positive-definite matrix. We denote its inverse and determinant by $g^{-1}\left(x,y\right)$ and $\left|g\left(x,y\right)\right|=\det g\left(x,y\right)$, respectively. These quantities are evaluated pointwise in the $\left(x,y\right)$-coordinates and define the intrinsic geometry of the surface $\mathcal{M}$ in this chart \cite{hsu_stochastic_2002,ikeda_stochastic_1981}.

\subsection{Confining potential and Riemannian gradient}

Dynamics are driven by a smooth confining potential on $\Omega$,
\begin{equation}
V\left(x\right)=\frac{k}{2}|x-c|^2,\qquad\qquad x=\left(x,y\right)^\top,\ \in\mathbb{R}^2,
\end{equation}
with $c=\left(0,0\right)$ and $k=2.8$. The Euclidean gradient is $\nabla V\left(x\right)=k\left(x-c\right)$. The corresponding Riemannian gradient, consistent with the metric $g$, is
\begin{equation}
\nabla_g V\left(x\right)=g\left(x\right)^{-1}\nabla V\left(x\right).
\end{equation}
This is the quantity used in the deterministic part of the drift.

\subsection{Riemannian Langevin diffusion in It\^{o} form}

We study an overdamped Langevin diffusion on $\Omega$ whose infinitesimal covariance is aligned with the inverse metric. Fix an inverse temperature parameter $\beta>0$. Define the diffusion tensor
\begin{equation}
a\left(x\right)=\frac{2}{\beta}\,g\left(x\right)^{-1}.
\end{equation}
To obtain a coordinate-invariant It\^o SDE with invariant measure proportional to
\[
\exp\!\left(-\beta V(x)\right)\sqrt{|g(x)|}\,dx
\]
(i.e., the Gibbs density with respect to the Riemannian volume element), the drift must include the standard It\^o correction term associated with the position-dependent diffusion \cite{risken_fokker-planck_1996,pavliotis_stochastic_2014}. In the implementation, the drift is written as
\begin{equation}
b\left(x\right)=-\nabla_g V\left(x\right)+\frac{1}{\beta}\,c\left(x\right),
\end{equation}
where $c\left(x\right)\in\mathbb{R}^2$ is the divergence-like correction
\begin{equation}
c^i\left(x\right)=\frac{1}{\sqrt{\left|g\left(x\right)\right|}}\ \partial_j\left(\sqrt{\left|g\left(x\right)\right|}\, g^{ij}\left(x\right)\right),
\end{equation}
with Einstein summation over $j\in\{1,2\}$ and $g^{ij}$ the components of $g^{-1}$. In the code, $c\left(x\right)$ is approximated by finite differences applied to the matrix field $\sqrt{\left|g\right|}\, g^{-1}$ on each coordinate direction, then divided by $\sqrt{\left|g\right|}$. No additional terms are introduced.

Let $W_t$ be a standard $\mathbb{R}^2$ Brownian motion. The continuous-time model underlying the numerical scheme is the It\^{o} SDE
\begin{equation}
dX_t=b\left(X_t\right)\,dt+\sqrt{a\left(X_t\right)}\,dW_t,\qquad\qquad a\left(x\right)=\frac{2}{\beta}g\left(x\right)^{-1},
\end{equation}
where $\sqrt{a\left(x\right)}$ is any matrix square root satisfying $\sqrt a\,\sqrt{a^{\top}}=a$ \cite{pavliotis_stochastic_2014}. The implementation constructs $\sqrt{g^{-1}\left(x\right)}$ by eigen-decomposition and uses
\begin{equation}
\sqrt{a\left(x\right)}=\sqrt{\frac{2}{\beta}}\ \sqrt{g^{-1}\left(x\right)}.
\end{equation}

\subsection{Discrete-time approximation (Euler--Maruyama)}

Trajectories are generated by an Euler--Maruyama discretization in the $\left(x,y\right)$-coordinates \cite{pavliotis_stochastic_2014}. Given a time step $\Delta t$, the update is
\begin{equation}
X_{n+1}=X_n+b\left(X_n\right)\,\Delta t+\sqrt{\frac{2}{\beta}}\ \sqrt{g\left(X_n\right)^{-1}}\ \Delta W_n,\qquad\qquad \Delta W_n\sim\mathcal{N}\left(0,\Delta t\, I_2\right).
\end{equation}
The matrix square root $\sqrt{g^{-1}}$ is computed via symmetric eigendecomposition with eigenvalues clipped to $\left[0,\infty\right)$ to ensure numerical stability.

\subsection{Domain restriction and boundary operators}

All dynamics are restricted to a rectangular domain $\Omega=\left[x_{min},x_{max}\right]\times\left[y_{min},y_{max}\right]$. After each Euler--Maruyama update, the state is mapped back into $\Omega$ using one of two boundary operators:
\begin{itemize}
\item Reflecting (Neumann-type) boundary: each coordinate is repeatedly mirror-reflected across the violated boundary until it lies in $\left[x_{min},x_{max}\right]$ (and analogously for $y$). This implements a specular reflection in coordinates and enforces confinement without absorption.
\item Periodic boundary: each coordinate is wrapped modulo the side length, i.e.,
\begin{equation}
x\mapsto x_{min}+\left(\left(x-x_{min}\right)\bmod L_x\right),\qquad\qquad L_x=x_{max}-x_{min},
\end{equation}
and similarly for $y$. This identifies opposite faces of the rectangle and produces a toroidal state space in coordinates.
\end{itemize}
These boundary choices are treated as experimental controls; the interior drift $b\left(x\right)$ and diffusion $a\left(x\right)$ are unchanged.

\subsection{Probability current and continuum entropy production}

Let $p\left(x,t\right)$ denote the probability density of $X_t$ with respect to Lebesgue measure in the $\left(x,y\right)$-coordinates. The associated Fokker--Planck equation can be written in conservative form \cite{risken_fokker-planck_1996},
\begin{equation}
\partial_t p\left(x,t\right)=-\nabla\cdot J\left(x,t\right),
\end{equation}
where the probability current $J:\Omega\times\operatorname{R}_+\rightarrow\mathbb{R}^2$ is
\begin{equation}
J\left(x,t\right)=b\left(x\right)\, p\left(x,t\right)-\frac{1}{2}\,\nabla\cdot\left(a\left(x\right)\, p\left(x,t\right)\right).
\end{equation}
Here the divergence of the matrix field $a p$ is taken row-wise:
\begin{equation}
\left[\nabla\cdot\left(ap\right)\right]_i=\sum_{j=1}^{2}{\partial_{x_j}\left(a_{ij}\left(x\right)\, p\left(x,t\right)\right)}.
\end{equation}
Given $J$, the (instantaneous) entropy production density used throughout is
\begin{equation}
\sigma\left(x,t\right)=\frac{2\, J\left(x,t\right)^\top a\left(x\right)^{-1}J\left(x,t\right)}{p\left(x,t\right)}.
\end{equation}
The corresponding continuum entropy production rate is the spatial integral
\begin{equation}
\dot{S_{\mathrm{cont}}}\left(t\right)=\int_{\Omega}\sigma\left(x,t\right)\,dx.
\end{equation}
In practice, we evaluate a steady-state approximation by estimating an empirical long-time density $p\left(x\right)$ and constructing the stationary current $J\left(x\right)=b\left(x\right)p\left(x\right)-\frac{1}{2}\nabla\cdot\left(a\left(x\right)p\left(x\right)\right)$, after which $\sigma\left(x\right)$ and $\int\sigma\left(x\right)\,dx$ are computed \cite{spinney_entropy_2012,van_den_broeck_three_2010}.

\subsection{Coarse-grained Markov entropy production under spatial binning}

To compare continuum irreversibility with a discrete coarse-graining, we induce a finite-state Markov chain by partitioning $\Omega$ into a uniform $n\times n$ grid of bins. Let the resulting discrete state space have size $K=n^2$, and let $S_n\left(X\right)\in\{1,\ldots,K\}$ denote the bin index of a point $X\in\Omega$ (undefined only for points outside $\Omega$, which do not occur after boundary mapping). From simulated trajectories at time step $\Delta t$, we estimate transition probabilities $P_{ij}\approx\mathbb{P}\left(S_n\left(X_{t+\Delta t}\right)=j\mid S_n\left(X_t\right)=i\right)$. Given a stationary distribution $\pi$ for $P$, the discrete-time entropy production rate used is the standard time-reversal asymmetry functional
\begin{equation}
\dot{S_{\mathrm{Markov}}}=\frac{1}{\Delta t}\sum_{i,j}{\pi_iP_{ij}}\log{\left(\frac{\pi_iP_{ij}}{\pi_jP_{ji}}\right)}.
\end{equation}
This object depends on the spatial resolution $n$; varying $n$ therefore provides a multiscale view of irreversibility under increasingly fine coarse-graining \cite{schnakenberg_network_1976,roldan_entropy_2012,esposito_stochastic_2012}.

\subsection{Summary of theoretical objects used in computation}

All subsequent numerical computations depend only on the tuple $\left(\Omega,g,V,\beta,\ \mathrm{boundary\ operator}\right)$ through the derived quantities
\begin{equation}
b\left(x\right)=-g^{-1}\left(x\right)\nabla V\left(x\right)+\frac{1}{\beta}\frac{1}{\sqrt{\left|g\left(x\right)\right|}}\partial_j\left(\sqrt{\left|g\left(x\right)\right|}\, g^{ij}\left(x\right)\right),\qquad a\left(x\right)=\frac{2}{\beta}g^{-1}\left(x\right),
\end{equation}
together with either (i) the continuum entropy production functional defined via $J=bp-\tfrac12\nabla\cdot\left(ap\right)$ and $\sigma=2J^\top a^{-1}J/p$, or (ii) the coarse-grained Markov entropy production functional defined via the bin-induced transition matrix $P$ and its stationary distribution $\pi$.

\section{Numerical Experiments}

We performed controlled numerical experiments with the explicit aim of quantifying entropy production under (i) different boundary conditions, (ii) different domain sizes, (iii) different temporal discretizations, and (iv) different observational coarse-grainings. Throughout, the inverse temperature was fixed to $\beta=4.5$, the confining potential was the quadratic $V\left(x\right)=\frac{k}{2}\lVert x-c\rVert^2$ with $k=2.8$ and $c=(0,0)$, and the initial condition was $X_0=(2.1,1.8)$. A single pseudo-random seed ($42$) was used for all configurations to ensure that changes across conditions reflect parameter changes rather than RNG variability.

\subsection{Domains and boundary conditions}

Two rectangular state-space domains were considered,
\begin{equation}
\Omega_{2.6}=\left[-2.6,2.6\right]\times\left[-2.6,2.6\right],\qquad
\Omega_{3.2}=\left[-3.2,3.2\right]\times\left[-3.2,3.2\right].
\end{equation}
For each $\Omega$, we evaluated two boundary regimes. In the reflecting regime, states were mapped back into $\Omega$ by repeated mirror reflection across violated faces, producing a Neumann-type reflection in coordinates. In the periodic regime, states were mapped into $\Omega$ by wrapping modulo the side lengths $L_x=x_{max}-x_{min}$ and $L_y=y_{max}-y_{min}$. These regimes were applied identically after every Euler--Maruyama update, so that the numerical state always remained in $\Omega$.

\subsection{Temporal discretization and sampling}

For each $\left(\Omega,\ \mathrm{boundary\ mode}\right)$ configuration, trajectories were simulated over a fixed physical horizon $H=20.0$ using time steps $\Delta t\in\{0.04,\ 0.02,\ 0.01\}$. The corresponding number of steps was $T=\left\lceil H/\Delta t\right\rceil$. The number of independent trajectories was scaled inversely with $\Delta t$ according to
\begin{equation}
R=\max\left(60,\ \operatorname{round}\left(R_{\mathrm{base}}\frac{\Delta t_{\mathrm{base}}}{\Delta t}\right)\right),
\end{equation}
with $R_{\mathrm{base}}=140$ and $\Delta t_{\mathrm{base}}=0.04$. This yields $R=140,\ 280,\ 560$ for $\Delta t=0.04,\ 0.02,\ 0.01$, respectively, matching the printed experiment summaries. Each simulation produced an array $\{X_t^{\left(r\right)}\}_{r=1..R,\ t=0..T}\subset\Omega$.

\subsection{Increment extraction and boundary-aware velocities}

To support density estimation, we constructed an increment dataset from the simulated paths. For each trajectory and time index, we recorded the starting position $Z=X_t$ together with an empirical velocity $Y=\Delta X_t/\Delta t$, where $\Delta X_t=X_{t+\Delta t}-X_t$. To reduce boundary-induced artifacts, increments were only retained if the starting position lay at least a fixed margin $\delta_{\mathrm{inc}}=0.35$ away from each face of $\Omega$. Under periodic boundaries, displacements were additionally converted to minimum-image form before dividing by $\Delta t$: for each coordinate, $\Delta x$ was mapped to the equivalent value in $\left(-\frac{L_x}{2},\frac{L_x}{2}\right]$ (and analogously for $\Delta y$). This prevents spurious large velocities caused purely by coordinate wrapping \cite{norris_markov_2009}.

When the increment dataset exceeded a prescribed cap ($1.2\times 10^5$ samples), a uniform random subsample (fixed subsampling seed $0$) was taken to keep downstream regression numerically stable and memory-bounded.

\subsection{Empirical steady-state density estimation on a fixed grid}

For each configuration and $\Delta t$, we estimated an empirical long-time density $p\left(x\right)$ on a uniform $64\times64$ grid over $\Omega$. The estimator proceeded in two steps. First, a 2D histogram $H$ of the increment starting positions $\{Z\}$ was computed over the grid-aligned binning of $\Omega$. Second, $H$ was smoothed by separable Gaussian convolution with bandwidth $h_p=0.30$ in physical units. Convolution padding was matched to the boundary regime: reflection padding for reflecting boundaries and wrap padding for periodic boundaries. The resulting smoothed field was normalized to integrate to one under the grid quadrature rule, i.e.,
\begin{equation}
\sum_{i,j} p_{ij}\,\Delta x\,\Delta y=1,
\end{equation}
where $\Delta x$ and $\Delta y$ are the uniform grid spacings implied by $\Omega$ and the $64\times64$ grid.

\subsection{Drift field on the grid}

Entropy production calculations require a drift field $b\left(x\right)$ on the same grid as $p\left(x\right)$. In the experiments reported here, the drift field was evaluated from the analytical model,
\begin{equation}
b\left(x\right)=-\nabla_gV\left(x\right)+\frac{1}{\beta}\,c\left(x\right),
\end{equation}
with $c\left(x\right)$ the coordinate correction term
\begin{equation}
c^i\left(x\right)=\left(1/\sqrt{\left|g\left(x\right)\right|}\right)\,\partial_j\left(\sqrt{\left|g\left(x\right)\right|}\, g^{ij}\left(x\right)\right),
\end{equation}
approximated numerically in the implementation by finite differences \cite{risken_fokker-planck_1996,pavliotis_stochastic_2014}.

\subsection{Continuum entropy production from the probability current}

Given the grid fields $p\left(x\right)$, $b\left(x\right)$, and the diffusion matrix $a\left(x\right)=\frac{2}{\beta}g^{-1}\left(x\right)$ evaluated on the same grid, we computed the probability current
\begin{equation}
J\left(x\right)=b\left(x\right)\,p\left(x\right)-\frac{1}{2}\,\nabla\cdot\left(a\left(x\right)p\left(x\right)\right),
\end{equation}
where the divergence of the matrix field $a p$ is understood componentwise:
\begin{equation}
\left[\nabla\cdot\left(ap\right)\right]_i=\sum_{j=1}^{2}{\partial_{x_j}\left(a_{ij}\left(x\right)p\left(x\right)\right)}.
\end{equation}
Spatial derivatives were approximated numerically by second-order centered differences in the interior. Under periodic boundaries, derivatives were computed via circular shifts. Under reflecting boundaries, one-sided differences were used at the outermost grid lines (a Neumann-like treatment consistent with reflection padding used in density smoothing) \cite{risken_fokker-planck_1996,pavliotis_stochastic_2014}.

The local entropy production density was then computed pointwise as
\begin{equation}
\sigma\left(x\right)=\frac{2\, J\left(x\right)^\top a\left(x\right)^{-1}J\left(x\right)}{p\left(x\right)},
\end{equation}
with a small floor applied to $p$ for numerical stability. A discretized approximation of the continuum entropy production rate was obtained by grid integration,
\begin{equation}
\dot{S_{\mathrm{cont}}}\approx\sum_{i,j}\sigma_{ij}\,\Delta x\,\Delta y,
\end{equation}
with an important correction under reflecting boundaries: to suppress boundary inflation artifacts arising from the discrete reflection rule and derivative approximations, the integral was restricted to an interior region excluding a fixed physical margin $\delta_{\mathrm{EP}}=0.40$ from each face of $\Omega$. Concretely, the mask was implemented by dropping the outermost $\left\lceil\delta_{EP}/\Delta x\right\rceil$ grid columns and $\left\lceil\delta_{EP}/\Delta y\right\rceil$ grid rows. Under periodic boundaries, no such exclusion was applied and the full domain was integrated \cite{spinney_entropy_2012,van_den_broeck_three_2010}.

\subsection{Coarse-grained Markov entropy production at multiple resolutions}

In parallel, we computed a discrete, coarse-grained entropy production rate from a spatial Markov chain induced by binning the trajectories. For each $\Delta t$, the domain $\Omega$ was partitioned into an $n\times n$ grid with $n\in\{24,32,48\}$, yielding $K=n^2$ discrete states. Transition counts $C_{ij}$ and occupancies $O_i=\sum_{j} C_{ij}$ were accumulated over all trajectories and time indices. Transition probabilities were estimated using an adaptive Dirichlet smoothing scheme: for each row $i$,
\begin{equation}
P_{ij}=\frac{C_{ij}+\alpha_i/K}{O_i+\alpha_i},\qquad\qquad
\alpha_i=\operatorname{clip}\left(\alpha_0\,\frac{K}{O_i+K},\ \alpha_{min},\ \alpha_{max}\right),
\end{equation}
with $\alpha_0=1.0$, $\alpha_{min}=10^{-3}$, and $\alpha_{max}=2.0$. This construction reduces variance in poorly sampled bins while converging to the empirical estimator as occupancy grows.

To avoid numerical pathologies from never-visited states, we restricted the chain to bins with occupancy $O_i\geq m$, with $m=5$, renormalizing transition rows after restriction. The stationary distribution $\pi$ of the restricted chain was estimated by power iteration \cite{norris_markov_2009}. The Markov entropy production rate was then computed as
\begin{equation}
\dot{S_{\mathrm{Markov}}}=\frac{1}{\Delta t}\sum_{ij}{\pi_iP_{ij}}\log\left(\frac{\pi_iP_{ij}}{\pi_jP_{ji}}\right),
\end{equation}
with small probability floors applied to ensure the logarithm is well-defined. We report $\dot{S_{\mathrm{Markov}}}$ for each bin resolution $n$, enabling a direct multiscale comparison between coarse-grained irreversibility and the continuum current-based estimate \cite{schnakenberg_network_1976,roldan_entropy_2012,esposito_stochastic_2012,polettini_effective_2017}.

\subsection{Recorded outputs}

For each $\left(\Omega,\ \mathrm{boundary\ mode},\ \Delta t\right)$ condition we recorded the continuum entropy production functional $\dot{S_{\mathrm{cont}}}$ and the Markov entropy production rates $\dot{S_{\mathrm{Markov}}}\left(n\right)$ for $n\in\{24,32,48\}$. For a single representative time step $\left(\Delta t=0.01\right)$ per configuration, we additionally retained diagnostic fields $p\left(x\right)$ and $\sigma\left(x\right)$ together with a subset of trajectories for visualization, to permit qualitative inspection of sampling coverage and spatial localization of entropy production.

\section{Results}

All experiments were conducted at inverse temperature $\beta=4.5$ with identical local dynamics; only global constraints were varied. For each step size $\Delta t\in\{0.04,\ 0.02,\ 0.01\}$, the number of trajectories was scaled as
\[
R\approx R_{\mathrm{base}}\left(\frac{\Delta t_{\mathrm{base}}}{\Delta t}\right),
\]
with a minimum of $60$, yielding $R=\{140,280,560\}$ and $T=\{500,1000,2000\}$ time steps, respectively. Continuum entropy production (EP) was computed using the model drift $b$ and the probability current evaluated on a $64\times64$ grid with histogram-based stationary density estimation and separable Gaussian smoothing. Markov EP was computed from bin-induced transition matrices at multiple spatial resolutions $n\in\{24,32,48\}$.

\subsection{Scale dependence of coarse-grained entropy production}

Coarse-grained entropy production exhibited systematic dependence on observation scale. Increasing spatial resolution produced monotonic increases in estimated irreversibility, whereas decreasing timestep produced monotonic decreases. These trends were consistent across domains and boundary modes (Tables~\ref{tab:markov_small}--\ref{tab:markov_large}). Because the estimator is defined from one-step transition statistics, its magnitude reflects both dynamical structure and the resolution at which trajectories are observed. The observed scaling therefore confirms that discrete entropy production is intrinsically coarse-graining dependent \cite{esposito_stochastic_2012,polettini_effective_2017}.

\begin{table}[h!]
\centering
\caption{Markov entropy production for domain $[-2.6,2.6]^2$. Estimated irreversibility increases with spatial resolution and decreases with finer temporal sampling.}
\label{tab:markov_small}
\begin{tabular}{c|ccc}
$\Delta t$ & 24 bins & 32 bins & 48 bins \\
\hline
0.04 & 0.8267 & 2.3537 & 8.7596 \\
0.02 & 0.4135 & 0.9312 & 3.4234 \\
0.01 & 0.1242 & 0.4128 & 1.2488
\end{tabular}
\end{table}

\begin{table}[h!]
\centering
\caption{Markov entropy production for domain $[-3.2,3.2]^2$. Resolution-dependent scaling persists across domain sizes, indicating that it is intrinsic to the estimator rather than domain-specific.}
\label{tab:markov_large}
\begin{tabular}{c|ccc}
$\Delta t$ & 24 bins & 32 bins & 48 bins \\
\hline
0.04 & 0.4546 & 1.1545 & 4.4517 \\
0.02 & 0.2126 & 0.5220 & 1.8179 \\
0.01 & 0.0476 & 0.1951 & 0.6266
\end{tabular}
\end{table}

\subsection{Sensitivity of continuum entropy production to global constraints}

In contrast to coarse-grained estimates, continuum entropy production varied strongly with boundary topology and domain size (Tables~\ref{tab:cont_small}--\ref{tab:cross}). Periodic domains consistently yielded larger values than reflecting domains, despite identical local dynamics. However, unlike earlier runs, the magnitude of this separation now depends on both timestep and domain size, indicating a coupled interaction between topology, geometry, and numerical resolution \cite{lebowitz_gallavotticohen-type_1999,spinney_entropy_2012,van_den_broeck_three_2010}.

\begin{table}[h!]
\centering
\caption{Continuum entropy production for domain $[-2.6,2.6]^2$. Boundary topology modulates entropy production, with periodic domains generally exhibiting higher values.}
\label{tab:cont_small}
\begin{tabular}{c|ccc}
Boundary & $\Delta t=0.04$ & $\Delta t=0.02$ & $\Delta t=0.01$ \\
\hline
Reflecting & 31.5319 & 4.4533 & 4.5685 \\
Periodic   & 62.9833 & 381.4358 & 73.9826
\end{tabular}
\end{table}

\begin{table}[h!]
\centering
\caption{Continuum entropy production for domain $[-3.2,3.2]^2$. Domain enlargement alters the magnitude and scaling of entropy production but preserves the topology-induced separation.}
\label{tab:cont_large}
\begin{tabular}{c|ccc}
Boundary & $\Delta t=0.04$ & $\Delta t=0.02$ & $\Delta t=0.01$ \\
\hline
Reflecting & 90.9782 & 74.0690 & 106.4949 \\
Periodic   & 240.8553 & 164.5243 & 216.6240
\end{tabular}
\end{table}

\begin{table}[h!]
\centering
\caption{Cross-domain comparison at $\Delta t=0.01$. Boundary effects interact with spatial extent, indicating coupled dependence on topology and geometry.}
\label{tab:cross}
\begin{tabular}{c|cc}
Domain & Reflecting & Periodic \\
\hline
$[-2.6,2.6]^2$ & 4.5685 & 73.9826 \\
$[-3.2,3.2]^2$ & 106.4949 & 216.6240
\end{tabular}
\end{table}

\subsection{Spatial structure of entropy production}

\begin{figure}[H]
\centering
\includegraphics[width=\linewidth]{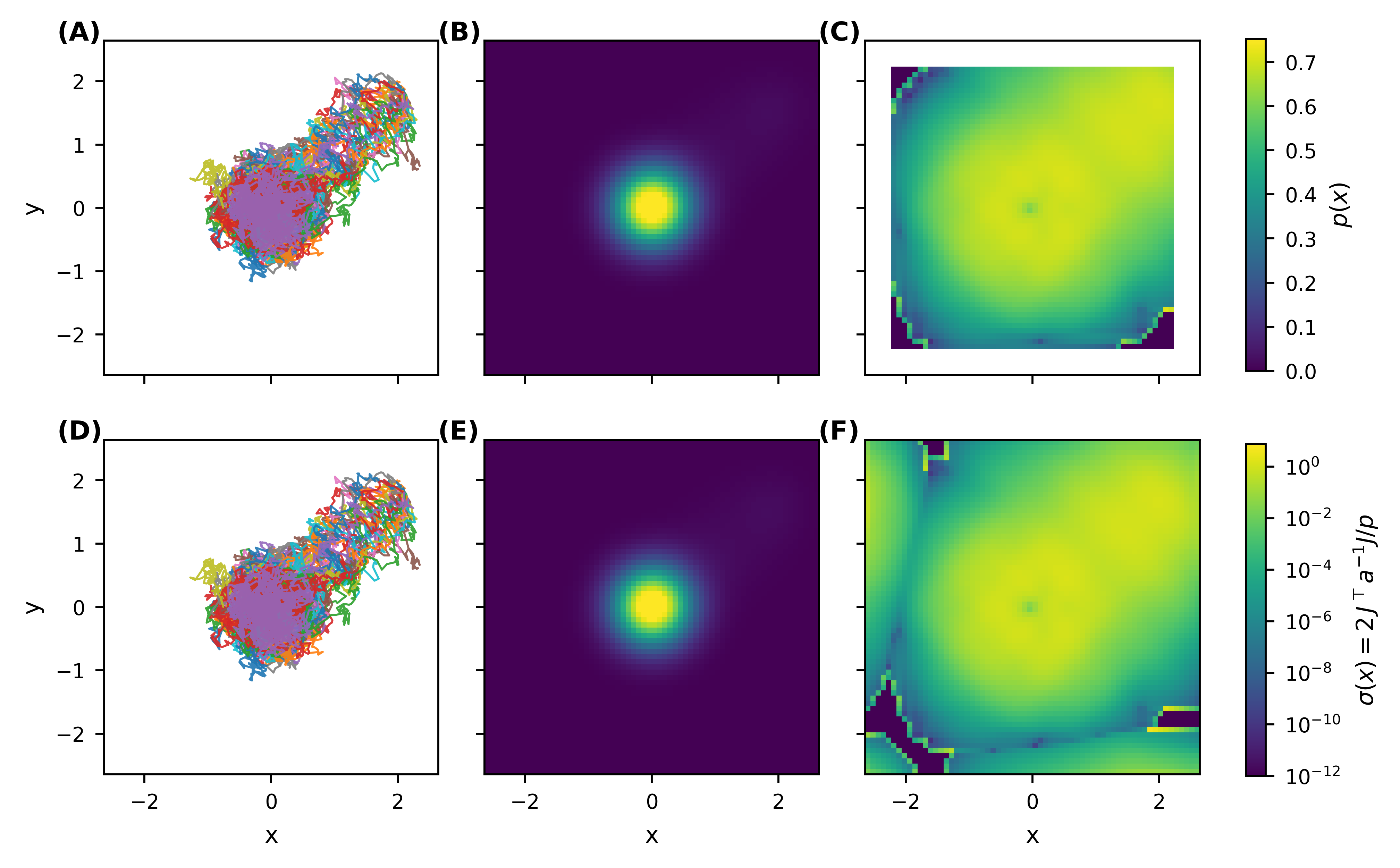}
\caption{\textbf{Spatial diagnostics for domain $[-2.6,2.6]^2$ at $\Delta t=0.01$.}
\textbf{Top row (A–C): reflecting boundaries.}
\textbf{Bottom row (D–F): periodic boundaries.}
(A,D) Sample trajectories.
(B,E) Estimated stationary density $p(x)$.
(C,F) Continuum entropy production density $\sigma(x)=2\,J^\top a^{-1}J/p$ (log scale).
Local stochastic dynamics are identical in both rows; only global boundary topology differs.
The stationary density remains smooth across conditions, whereas $\sigma(x)$ exhibits structured heterogeneity characteristic of irreversible probability transport.}
\label{fig:fields_small}
\end{figure}

\begin{figure}[H]
\centering
\includegraphics[width=\linewidth]{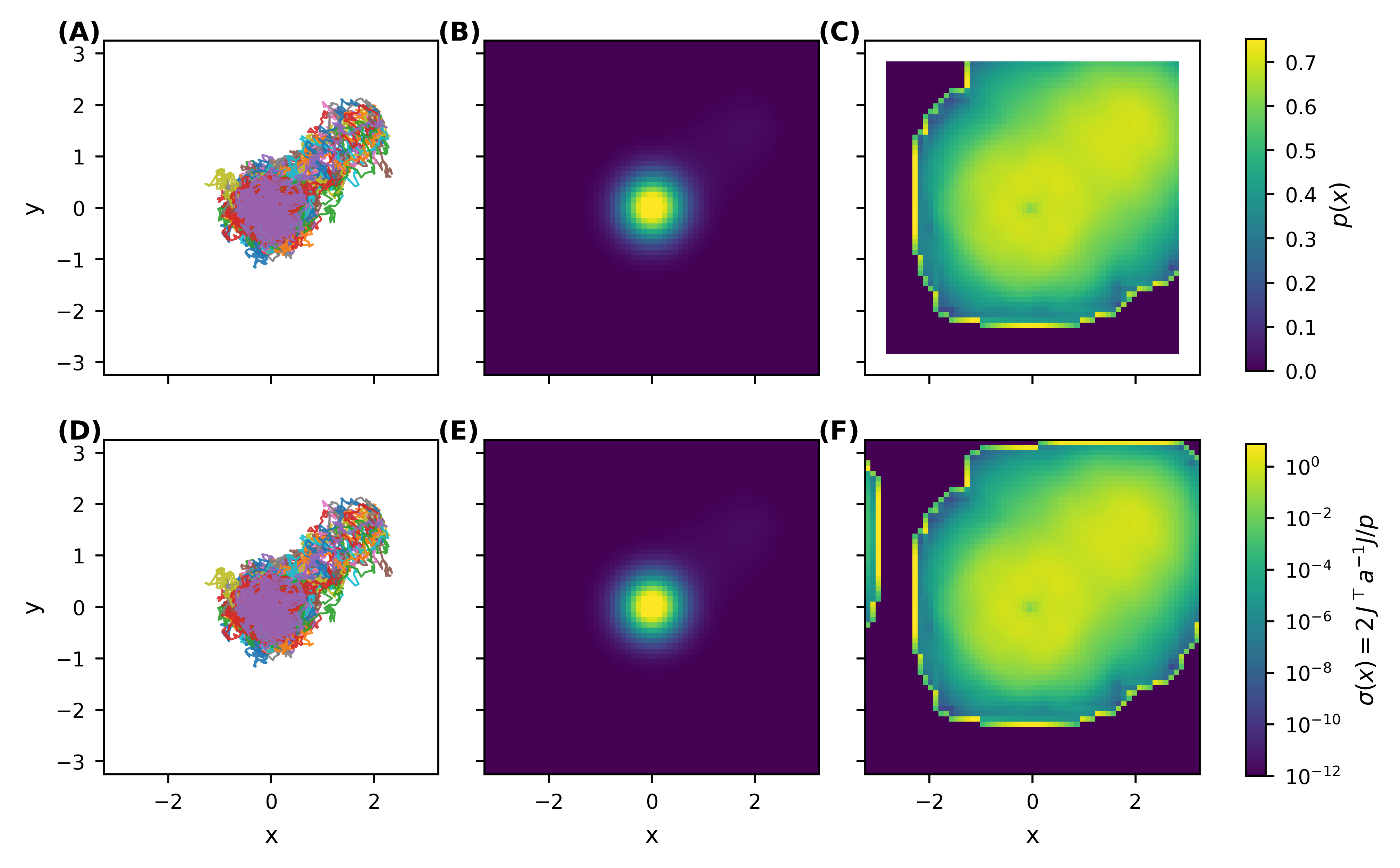}
\caption{\textbf{Spatial diagnostics for domain $[-3.2,3.2]^2$ at $\Delta t=0.01$.}
\textbf{Top row (A–C): reflecting boundaries.}
\textbf{Bottom row (D–F): periodic boundaries.}
(A,D) Sample trajectories.
(B,E) Estimated stationary density $p(x)$.
(C,F) Continuum entropy production density $\sigma(x)=2\,J^\top a^{-1}J/p$ (log scale).
Increasing domain size alters the magnitude and geometry of entropy production but preserves the topology-dependent separation between rows.
Regions of elevated $\sigma(x)$ coincide with sustained probability transport rather than density maxima.}
\label{fig:fields_large}
\end{figure}

Spatially resolved diagnostics reveal that elevated entropy production aligns with regions of sustained probability transport rather than regions of maximal stationary density. In both domain sizes, the estimated stationary density is broadly smooth, whereas the entropy production density exhibits structured heterogeneity that reflects irreversibility carried by probability current rather than stochastic dispersion (Figures~\ref{fig:fields_small}--\ref{fig:fields_large}). These fields support the interpretation of entropy production as a measure of irreversible flow organization induced by global constraints \cite{gingrich_dissipation_2016,nardini_entropy_2017}.

\FloatBarrier

\subsection{Empirical synthesis}

Across all experimental configurations, five consistent empirical regularities emerged:

\begin{enumerate}
\item Coarse-grained entropy production increases monotonically with spatial resolution.
\item Coarse-grained entropy production decreases monotonically with finer temporal discretization.
\item Continuum entropy production is strongly modulated by boundary topology.
\item Periodic domains generally produce larger entropy production than reflecting domains despite identical local stochastic structure.
\item Domain size modifies, but does not eliminate, the influence of topology on irreversibility.
\end{enumerate}

Taken together, these results demonstrate that the dominant determinant of entropy production in the system is constraint-induced probability transport rather than local stochastic variability.


\section{Discussion}

The objective of this study was to interrogate a widely held intuition: that entropy production functions as a quantitative proxy for microscopic disorder or environmental roughness in stochastic systems. By constructing a controlled computational setting in which local geometry, stochastic forcing, and drift structure were held fixed while global constraints were selectively varied, we were able to isolate what actually governs entropy production across scales, resolutions, and estimators. This separation of local dynamics from global admissibility conditions allows the role of constraints themselves to be examined directly rather than inferred indirectly \cite{seifert_stochastic_2012,spinney_entropy_2012}.

Across all experiments, a consistent pattern emerged. Entropy production was not primarily determined by local stochastic variability or geometric irregularity, but instead by whether the system’s state space permitted sustained probability currents. In other words, irreversibility was dictated not by how noisy the dynamics were, but by whether the geometry and topology of the domain allowed motion to organize into directional transport \cite{lebowitz_gallavotticohen-type_1999,van_den_broeck_three_2010}.

The clearest evidence arises from boundary topology. For identical local dynamics, periodic domains consistently produced larger continuum entropy production than reflecting domains. Because the noise amplitude, surface geometry, and drift field were unchanged, this separation cannot be attributed to microscopic disorder. Rather, it reflects a structural distinction: periodic boundaries allow probability mass to circulate, whereas reflecting boundaries interrupt such currents and promote local equilibration. Entropy production therefore scales with the system’s capacity for directional transport, not with the unpredictability of individual trajectories. The magnitude of this separation varies with timestep and domain size, indicating that topology interacts with discretization and geometry rather than acting in isolation \cite{gaspard_time-reversal_2012,bertini_macroscopic_2015}.

Temporal resolution provides a second, independent perspective. As the timestep was refined, coarse-grained entropy production estimates systematically decreased. This behavior indicates that part of the irreversibility inferred at coarse resolution arises from temporal aggregation, which masks small-scale reversibility. When observations resolve finer temporal structure, transitions appear more symmetric and estimated entropy production declines. Such behavior would not occur if entropy production directly measured intrinsic noise intensity, which remained fixed throughout all experiments \cite{esposito_stochastic_2012,polettini_effective_2017}.

Spatial resolution reveals a complementary effect. Increasing the number of bins in the coarse-grained Markov estimator increased measured entropy production across all configurations. This scaling demonstrates that insufficient spatial resolution obscures directional probability fluxes: transitions that encode organized currents are averaged together and appear artificially reversible. Consequently, the magnitude of discrete entropy production is determined jointly by the system’s dynamics and the observer’s level of coarse-graining. What appears irreversible at one scale may appear nearly reversible at another \cite{schnakenberg_network_1976,roldan_entropy_2012}.

Taken together, these findings show that entropy production fundamentally quantifies the extent to which a system violates detailed balance through structured probability transport. Stochastic forcing generates motion, but does not by itself produce large entropy production. Instead, substantial entropy production requires sustained flux in state space, and such flux is only possible when global constraints permit it. Topology, boundary conditions, and admissible trajectory geometry therefore play a determining role \cite{maes_active_2014,bertini_macroscopic_2015}. The present study considers non-interacting trajectories; extension to interacting or high-dimensional systems remains an open direction.

This distinction resolves an important conceptual ambiguity. Randomness alone produces dispersion, but dispersion is not synonymous with irreversibility. Irreversibility requires directional structure—organized currents that break time-reversal symmetry. Entropy production measures precisely this asymmetry. It is therefore best understood not as a scalar index of disorder, but as a quantitative signature of nonequilibrium flow organization \cite{seifert_entropy_2005,seifert_stochastic_2012}.

Finally, these findings clarify the geometric dimension of stochastic irreversibility. The dynamics considered here evolve on a curved manifold with position-dependent metric, so both drift and diffusion tensors vary spatially. Yet altering boundary topology produced more pronounced and systematic effects than modifying resolution alone, indicating that global structural constraints exert dominant influence over nonequilibrium behavior. Noise supplies fluctuations, but constraints determine whether those fluctuations can assemble into persistent transport \cite{hsu_stochastic_2002,ikeda_stochastic_1981}.


\section{Conclusion}

The central empirical result is that entropy production changes systematically when global constraints are altered while all local dynamical ingredients remain fixed. Periodic domains, which permit sustained circulation, consistently exhibit higher entropy production than reflecting domains, which suppress global transport, demonstrating that entropy production is governed primarily by the geometry of allowable trajectories rather than by the magnitude of stochastic fluctuations; more broadly, the findings support a reinterpretation of entropy production as a diagnostic of constrained flow structure in state space, where spatial maps of entropy production reveal where probability mass is actively transported and thereby expose hidden dynamical constraints, circulation pathways, and irreversible mechanisms that cannot be detected from trajectory variance alone, aligning with current-based formulations of irreversibility \cite{lebowitz_gallavotticohen-type_1999,spinney_entropy_2012,van_den_broeck_three_2010} and with broader links between dissipation, nonequilibrium structure, and information-theoretic asymmetry \cite{kawai_dissipation_2007,parrondo_thermodynamics_2015,cover_elements_2005}.

\section*{Code Availability}

All code required to reproduce the results of this study is openly available at
\url{https://github.com/patrickromanescu/entropy-production}.
The repository includes simulation scripts, analysis routines, and figure-generation code corresponding to all results reported in the manuscript. The code is distributed under the MIT License.

\bibliography{references}

\end{document}